\documentclass[
    ,final            
  ]
  {aipproc}

\layoutstyle{8x11double}

\newcommand {\beq}{\begin{eqnarray}}
\newcommand {\eeq}{\end{eqnarray}}
\newcommand {\non}{\nonumber\\}

\newcommand {\1}[1]{\frac{1}{#1}}

\newcommand{\hs}[1]{\hspace{#1 mm}}

\newcommand {\Nf}{N_{\rm F}} 
\newcommand {\Nc}{N_{\rm C}} 

\newcommand {\Tr}{\rm Tr}
\def\D{\mathcal{D}}

\begin{document}

\title{D-brane Configurations for Domain Walls and Their Webs}



\author{Minoru Eto}{
  address={Department of Physics, Tokyo Institute of 
Technology, Tokyo 152-8551, JAPAN}
}

\author{Youichi Isozumi}{
  address={Department of Physics, Tokyo Institute of 
Technology, Tokyo 152-8551, JAPAN}
}

\author{Muneto Nitta}{
  address={Department of Physics, Tokyo Institute of 
Technology, Tokyo 152-8551, JAPAN}
}

\author{Keisuke Ohashi}{
  address={Department of Physics, Tokyo Institute of 
Technology, Tokyo 152-8551, JAPAN}
}

\author{Kazutoshi Ohta}{
  address={
Theoretical Physics Laboratory\\
The Institute of Physical and Chemical Research (RIKEN)\\
2-1 Hirosawa, Wako, Saitama 351-0198, JAPAN}
}

\author{Norisuke Sakai}{
  address={Department of Physics, Tokyo Institute of 
Technology, Tokyo 152-8551, JAPAN}
}

\begin{abstract}

Supersymmetric $U(\Nc)$ gauge theory 
with $\Nf$ massive hypermultiplets 
in the fundamental representation 
admits various BPS solitons like 
domain walls and their webs. 
In the first part 
we show as a review of the previous paper \cite{Eto:2004vy} 
that domain walls are realized as kinky fractional D$3$-branes 
interpolating between separated D$7$-branes.
In the second part we discuss brane configurations 
for domain wall webs. 
This is a contribution to the conference based on the talk 
given by MN.

\end{abstract}

\maketitle


\section{Introduction}\label{INTRO}

Domain walls (or kinks) are  
important solitons in various subjects of physics 
like high energy physics, 
cosmology and condensed matter physics. 
They are BPS states in supersymmetric gauge theories.  
Like other BPS solitons their solutions admit 
the moduli space. 
Recently multiple domain wall solutions 
have been obtained 
and their moduli space 
has been completely determined 
in non-Abelian $U(\Nc)$ gauge theory 
coupled with the fundamental Higgs fields \cite{INOS1},
generalizing the $U(1)$ cases \cite{U(1)}. 
Soliton dynamics can be often well understood 
when solitons are embedded in 
D-brane configuration in 
string theory~\cite{Giveon:1998sr}. 
Domain wall solutions in \cite{INOS1} 
has been realized in \cite{Eto:2004vy} 
as a kinky brane configuration, 
first suggested in \cite{LT} for the $U(1)$ case. 

Since then there have been extensive developments 
about domain walls. 
1/4 BPS configurations of vortices 
stretched between multiple walls has been found 
in \cite{INOS3}, 
generalizing vortices ending on a single wall~\cite{wv}. 
The negative monopole charge firstly found in 
\cite{INOS3}, 
now called a boojum, 
has been further studied in \cite{boojum}. 
In \cite{Eto:2005wf} domain walls in 
$U(1)^2$ gauge theory has been investigated, 
where attractive and repulsive force 
between walls has been found.   
Relations between the moduli space of 
domain walls and one of monopoles has been 
clarified \cite{Hanany:2005bq}. 
Relations to flux tube has been discussed \cite{Bolognesi:2005zr}.
Domain walls serve a set up to prove 
a relation between 
Skyrmion and instantons \cite{Eto:2005cc}. 
Finally, it has been found that 
several domain walls with different angles in 
the real space 
can make a 1/4 BPS web~\cite{webs}. 
In the present talk 
the kinky D-brane configuration for 
domain walls \cite{Eto:2004vy} is reviewed 
and then is generalized to their webs.

\section{Domain Walls} \label{NAW}
In this section we discuss
domain walls~\cite{INOS1}. 
For notation see the original papers. 
The vector multiplet contains 
a gauge field $W_M$ 
and a complex scalar field $\Sigma = \Sigma_1 + i \Sigma_2$  
in the adjoint representation of the $U(N_{\rm C})$ gauge group 
whereas the hypermultiplets contain Higgs fields of  
two $\Nc \times \Nf$ matrices $(H^i)^{rA} \equiv (H^{irA})$ 
with $SU(2)_R$ $i=1,2$,  
color $r=1, \cdots, N_{\rm C}$ and flavor 
$A=1, \cdots, N_{\rm F}$ indices.  
The bosonic Lagrangian is  
(summation over repeated index $\alpha=1,2$ is implied 
in the following)
\beq
&& {\cal L} 
=
-\frac{1}{2g^2}{\rm Tr}\left( F_{MN}F^{MN}\right)
+ 
\frac{1}{g^2}{\rm Tr} 
\left({\cal D}_M \Sigma_{\alpha} {\cal D}^M \Sigma_{\alpha} \right) \non
&& \hs{5} +{\rm Tr}
\left[ {\cal D}^M H^i ({\cal D}_M H^i)^\dagger \right] -V,
  \label{Lagrangian}
\eeq
with $g$ the gauge coupling constant and 
$V$ the potential: 
$V= \frac{g^2}{4}{\rm Tr}
\left[
\left(
H^{1}  H^{1\dagger}  - H^{2} H^{2\dagger} 
- c\mathbf{1}_{N_{\rm C}}
\right)^2
+4H^2H^{1\dagger} H^1H^{2\dagger} 
\right]$ 
$+{\rm Tr}\left[
 (\Sigma_{\alpha} H^i - H^i M_{\alpha}) 
 (\Sigma_{\alpha} H^i - H^i M_{\alpha})^\dagger 
- \frac{1}{g^2}\left[\Sigma_1,\Sigma_2\right]^2
 \right]$, 
with a triplet of the Fayet-Iliopoulos (FI) 
parameters chosen to the third direction as $(0,0,c)$,   
and with the mass matrix defined by 
$(M_{\alpha})^A{}_B\equiv (m_A,n_A)\delta ^A{}_B$. 
We consider non-degenerate real masses 
($n_A=0$) with ordering 
$m_{A+1} < m_{A}$ for domain walls, 
but complex masses $m_A + i n_A$ for their webs. 
In the massless case 
the moduli space of vacua 
becomes the Higgs branch, 
the cotangent bundle over the complex Grassmann manifold
${\cal M}^{M=0}_{\rm vacua} 
  \simeq T^* G_{\Nf,\Nc} 
  = T^* \left[{SU(\Nf) \over SU(\Nc) 
   \times SU(\tilde \Nc) \times U(1)} \right]$
with $\tilde \Nc \equiv \Nf - \Nc$~\cite{ANS}. 
Turning on masses for hypermultiplets, 
most points on 
${\cal M}^{M=0}_{\rm vacua}$ are lifted leaving 
the discrete SUSY vacua given by 
$H^{1rA}=\sqrt{c}\,\delta ^{A_r}{}_A,\quad H^{2rA}=0, \quad
\Sigma ={\rm diag.}(m_{A_1} + i n_{A_1},\cdots,\,
           m_{A_{N_{\rm C}}} + i n_{A_{N_{\rm C}}})$, 
denoted by
$\langle A_1,\cdots, A_{\Nc}\rangle$,  
and therefore the number of SUSY vacua is 
${}_{\Nf} C_{\Nc} = {\Nf ! \over \Nc ! \tilde \Nc!}$.

The BPS equations for 
1/2 BPS domain walls are obtained as \cite{INOS1} 
($H^1 \equiv H, H^2 =0$)
\begin{eqnarray}
 \D_y \Sigma =  
 {g^2\over 2}\left(c{\bf 1}_{N_{\rm C}}-H H^\dagger\right), 
 \quad \D_y H = -\Sigma H + H M, \label{BPSeqs}
\end{eqnarray}
with $y$ the codimension of walls. 
The tension of multiple BPS walls, 
interpolating between a vacuum 
$\langle A_1,\cdots, A_{\Nc}\rangle$ at $y \to + \infty$ 
and a vacuum 
$\langle B_1,\cdots, B_{\Nc}\rangle$ at $y \to - \infty$, 
is obtained as
\begin{eqnarray}
 Z = c \left[{\rm Tr}\Sigma \right]^{y = +\infty}_{y = -\infty}
\!\!=\!c \left(\sum_{r=1}^{N_{\rm C}}m_{A_r}
-\sum_{r=1}^{N_{\rm C}}m_{B_r}\right) . 
\label{eq:tension} 
\end{eqnarray}
See the original papers 
\cite{INOS1}
for how to solve Eqs.~(\ref{BPSeqs}). 

We now realize our theory on D-branes 
in string theory as follows.
Let us consider parallel  
$\Nf$ D$7$-branes 
in type IIB string theory and 
divide four spatial directions of their world volume 
by ${\bf Z}_2$ to form 
the orbifold ${\bf C}^2 / {\bf Z}_2$.
The orbifold singularity is blown up to the Eguchi-Hanson space 
by $S^2$ with the area 
\beq
 A = c g_s l_s^{4} = {c \over \tau_3}  \label{S2area}
\eeq
with $g_s$ the string coupling, 
$l_s=\sqrt {\alpha'}$ the string length and 
$\tau_{3} = 1/ g_s l_s^{4}$ 
the D$3$-brane tension.  
We then realize our theory on 
$\Nc$ fractional D$3$-branes, 
that is, D$5$-branes wrapping around $S^2$. 
Then the configuration becomes 
\beq
 \mbox{$\Nc$ frac. D$3$:} && 0123     \non
 \mbox{$\Nf$ D$7$:} && 01234567 \non
 \mbox{${\bf C}^2/{\bf Z}_2$ ALE:}  && \hs{7} 4567 .
\eeq 
A string connecting D$3$-branes provides the gauge multiplets 
whereas a string connecting D$3$-branes and D$7$-branes 
provides the hypermultiplets in the fundamental representation. 
The gauge coupling constant $g$ of the gauge theory realized on the 
D$3$-brane is 
\beq
 \1{g^2} 
   = {b \over \, g_s } \label{g^2}
\eeq
with 
$b$ the $B$-field flux integrated over the $S^2$, 
$b \sim A B_{ij}$. 
The positions of the D$7$-branes in the $x^8$-$x^9$ plane gives 
the complex masses for the fundamental hypermultiplets 
whereas the positions of the D$3$-branes in the $x^8$-$x^9$ plane 
is determined by the VEV of $\Sigma$: 
\beq 
&& (x^8,x^9)|_{{\rm D}7} 
       = l_s^2 (m,n)  , \non
&& (x^8,x^9)|_{{\rm D}3} 
   = l_s^2  (\Sigma^1(x^1,x^2),\Sigma^2(x^1,x^2)) . 
\eeq

Taking a T-duality along the $x^4$-direction 
the ALE geometry is mapped to two NS$5$-branes 
separated in the $x^4$-direction. 
The configuration becomes 
the Hanany-Witten type brane configuration \cite{HW}
\beq
 \mbox{$\Nc$ D$4$:} && 01234     \non
 \mbox{$\Nf$ D$6$:} && 0123 \hs{2} 567 \non
 \mbox{$2$ NS$5$:}  && 0123 \hs{8.5} 89 . 
  \label{HWset1}
\eeq
The relations between physical quantities and 
the positions of branes are 
summarized as follows: 
\beq 
 && (x^8,x^9)|_{{\rm D}4} 
        = l_s^2 (\Sigma^1(x^1,x^2),\Sigma^2(x^1,x^2)), \non
 && (x^8,x^9)|_{{\rm D}6} = l_s^2 (m,n), 
    \quad \Delta x^4|_{{\rm NS}5} = {g_s l_s \over g^2}, \non
 && (\Delta x^5, \Delta x^6, \Delta x^7) |_{{\rm NS}5}
  = g_s l_s^3 (0,0,c).  \label{quantities}
\eeq

As vacuum states in the brane picture 
before taking T-duality, 
any D$3$-brane 
must lie in a D$7$-brane.  
At most one D$3$-brane  
can lie in each D$7$-brane 
(because of s-rule \cite{HW})
so that the vacuum 
$\langle A_1,\cdots, A_{\Nc}\rangle$ 
is realized with $A_r$ denoting 
positions of D$3$-branes, 
and therefore the number of 
vacua is ${}_{\Nf} C_{\Nc}$.  

\begin{figure}[thb]
\includegraphics[width=6cm,clip]{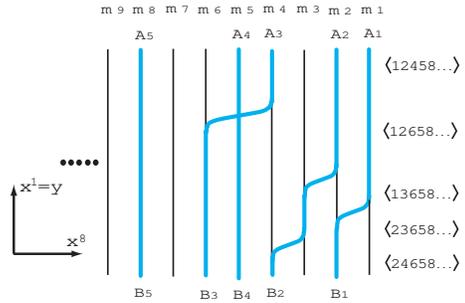}
\caption{\small
Multiple non-Abelian walls as kinky D-branes. 
}
\label{fig9}
\end{figure}

For domain wall states, we 
consider real mass $n=0$ and $\Sigma_2=0$, 
and $\Sigma_1$ depends on 
one coordinate $y \equiv x^1$. 
All D$3$-branes lie in 
some $\Nc$ D$7$-branes in the limit 
$y \to + \infty$, 
giving $\langle A_1,\cdots, A_{\Nc}\rangle$,
but lie in another set of D$7$-branes 
in the opposite limit $y \to - \infty$,  
giving another vacuum 
$\langle B_1,\cdots, B_{\Nc}\rangle$. 
The $\Nc$ D$3$-branes exhibit kinks somewhere 
in the $y$-coordinate  
as illustrated in Fig.~\ref{fig9}. 
Here we labeled $B_r$ such that 
the $A_r$-th brane at $y \to + \infty$ goes to the $B_r$-th brane 
at $y \to - \infty$. 
If we separate adjacent walls far enough 
the configuration between these walls 
approach a vacuum  as illustrated in Fig.~\ref{fig9}. 

\begin{figure}[thb]
\begin{tabular}{cc}
\includegraphics[width=3cm,clip]{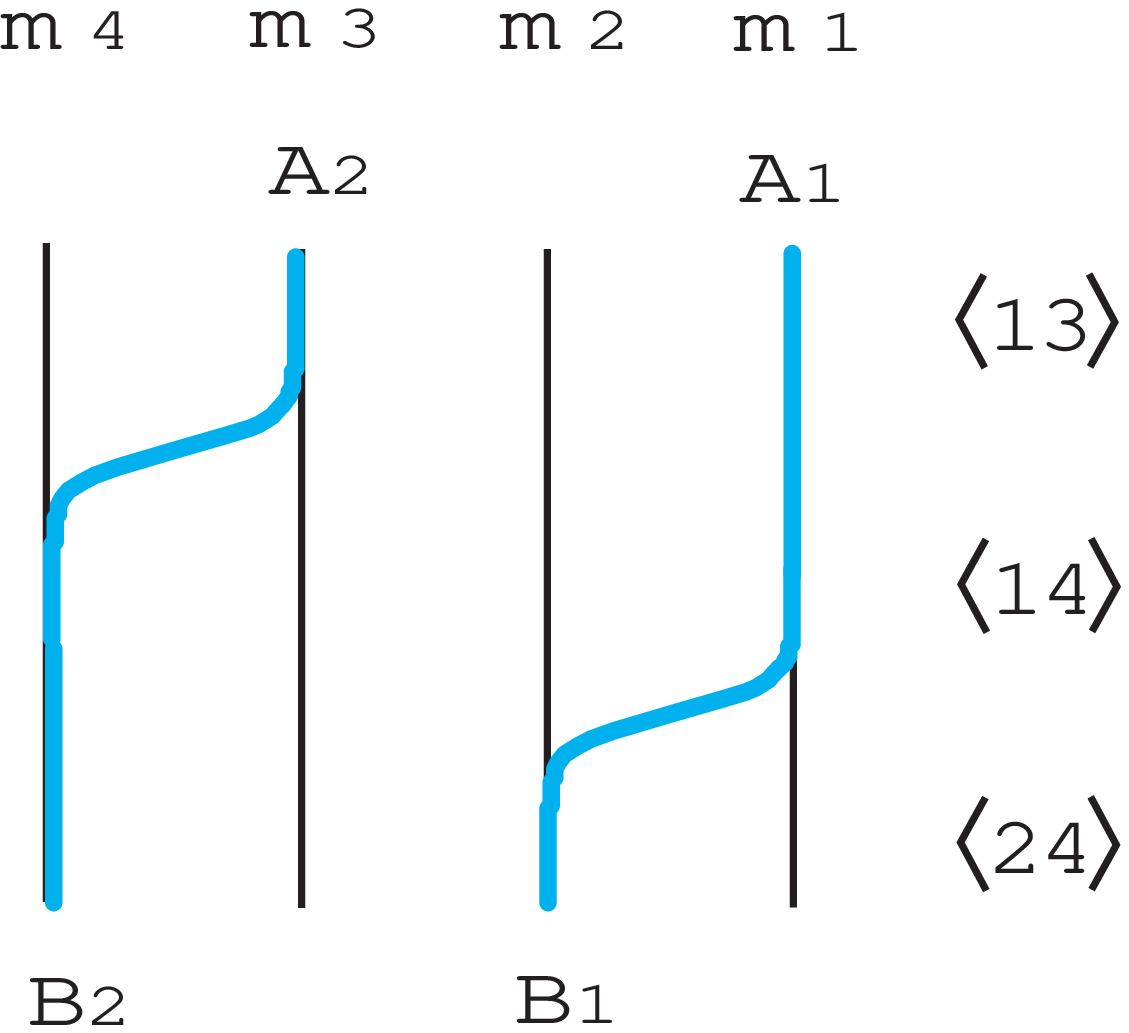}
 & 
\includegraphics[width=3cm,clip]{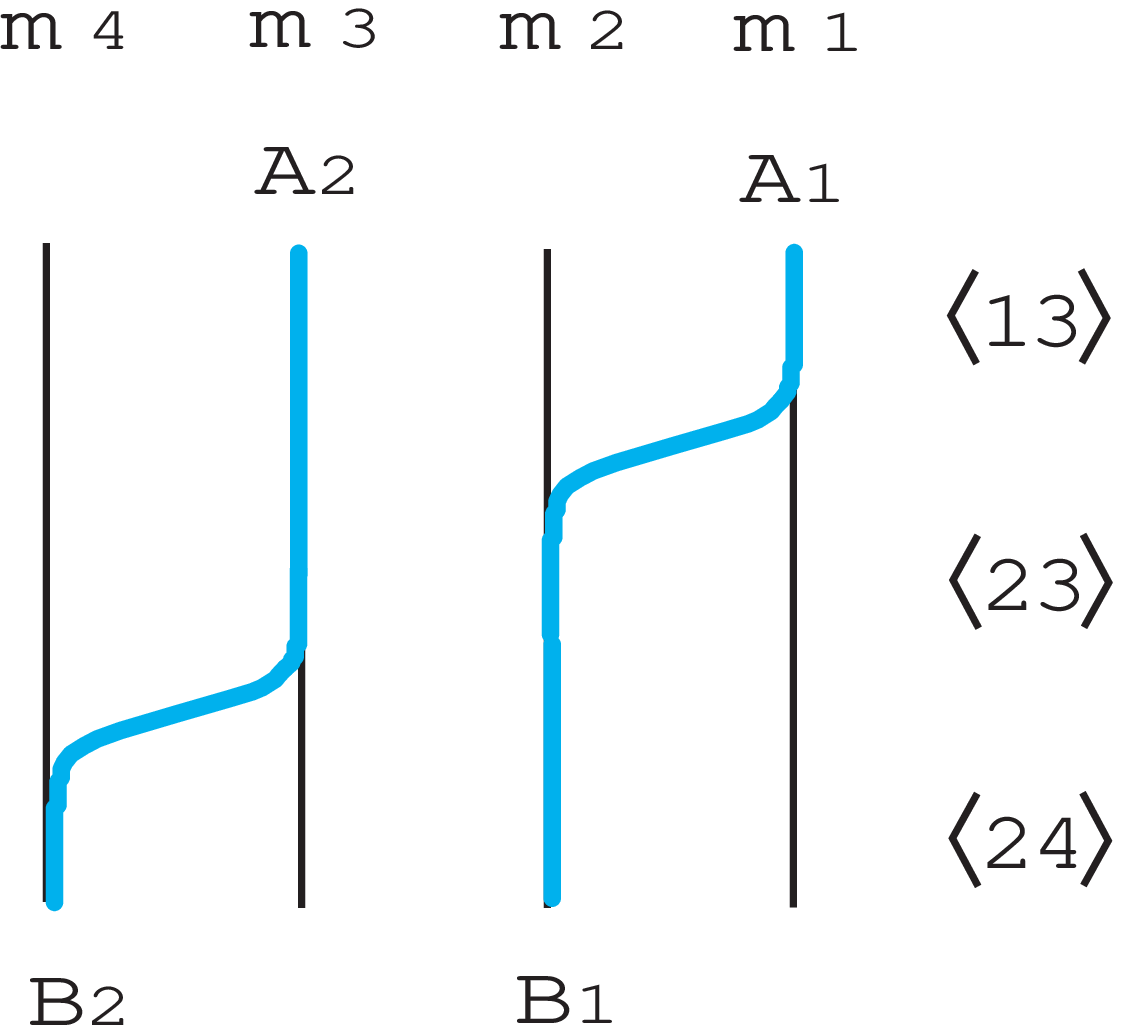} 
\end{tabular}
\caption{\small
Penetrable walls. 
}
\label{fig16}
\end{figure}

\begin{figure}[thb]
\begin{tabular}{cc}
\includegraphics[width=2.8cm,clip]{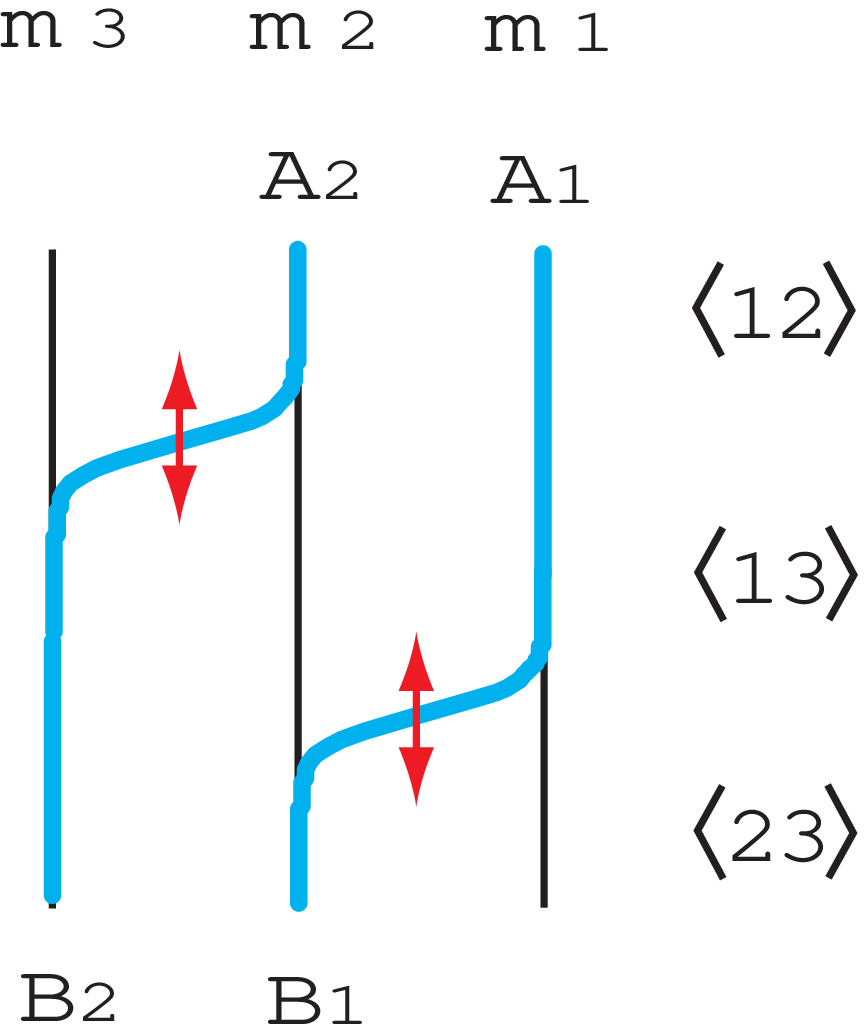}
&
\includegraphics[width=2.8cm,clip]{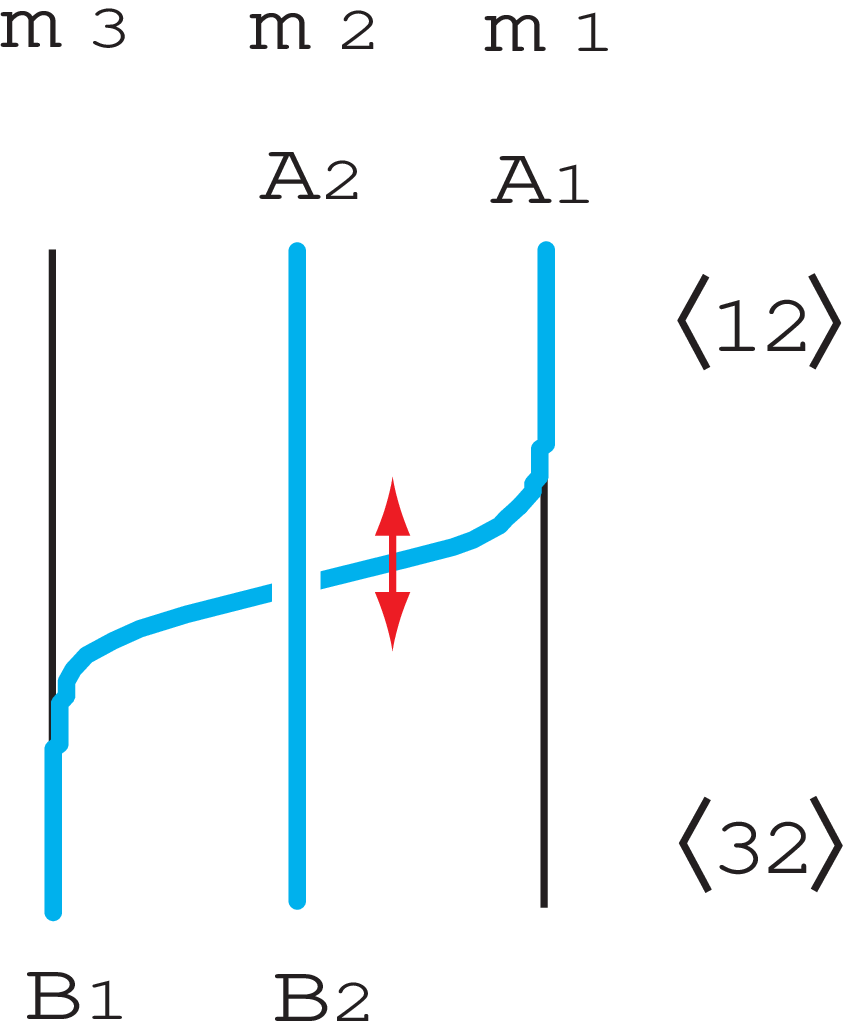}
\end{tabular}
\caption{\small
Impenetrable walls
}
\label{fig13}
\end{figure}

Using quantities given in
(\ref{S2area}) and (\ref{g^2}), 
or (\ref{quantities}) in T-dualized picture, 
the tension formula 
(\ref{eq:tension}) of walls 
is correctly reproduced by 
the brane tension around kinky regions.
These configuration clarify 
dynamics of domain walls easily. 
In non-Abelian gauge theory 
two domain walls can penetrate each other 
if they are made of separated 
D$3$-branes like Fig.~\ref{fig16}
but they cannot  
if they are made of adjacent 
D$3$-branes like Fig.~\ref{fig13}. 
In the latter case, 
reconnection of D$3$-branes 
occur in the limit that two walls are compressed.


\section{Domain Wall Webs}

The BPS equations for domain wall webs can 
be obtained as \cite{webs} ($H^1 \equiv H, H^2 =0$)
\beq
&&F_{12} = i \left[\Sigma_1,\Sigma_2\right], \quad 
 \D_1\Sigma_2 = \D_2\Sigma_1,  \non
&&  \D_1\Sigma_1 + \D_2\Sigma_2 
 = {g^2\over 2}\left(c{\bf 1}_{N_{\rm C}}-H H^\dagger 
   \right),  \label{bps_eq1} \\
&&\D_1 H = HM_1 - \Sigma_1 H,\quad
  \D_2 H = HM_2 - \Sigma_2 H. \nonumber
\eeq
with $x^1$ and $x^2$ the codimensions.
The energy density is 
\beq
{\cal E} 
 = {\cal Y}+ {\cal Z}_1 + {\cal Z}_2 
+  
  \partial_\alpha J_\alpha,
\label{energy_comp}
\eeq
where each charge density is given by
\beq
{\cal Y} \equiv \frac{2}{g^2}
\partial_\alpha\Tr\left(\epsilon^{\alpha\beta}
\Sigma_2\D_\beta\Sigma_1\right),\;
{\cal Z}_{\alpha} 
 \equiv c \partial_{\alpha} \Tr \Sigma_{\alpha} 
 \small{(\mbox {no sum})}
\label{cetral charge}
\eeq 
with $J_{\alpha}$ terms not contributing to the tension 
under the space integration.
Wall tensions and the junction charge are defined by 
$Z_{\alpha} \equiv \int dx^{\alpha} {\cal Z}_{\alpha}$ 
and 
$Y \equiv \int dx^1 dx^2 {\cal Y}$, respectively. 
See the original papers \cite{webs}
for solutions to (\ref{bps_eq1}) and 
their properties. 
In the following 
we consider only $U(1)$ gauge group 
for simplicity.

\begin{figure}[thb]
\begin{tabular}{cc}
 \includegraphics[width=3cm,clip]{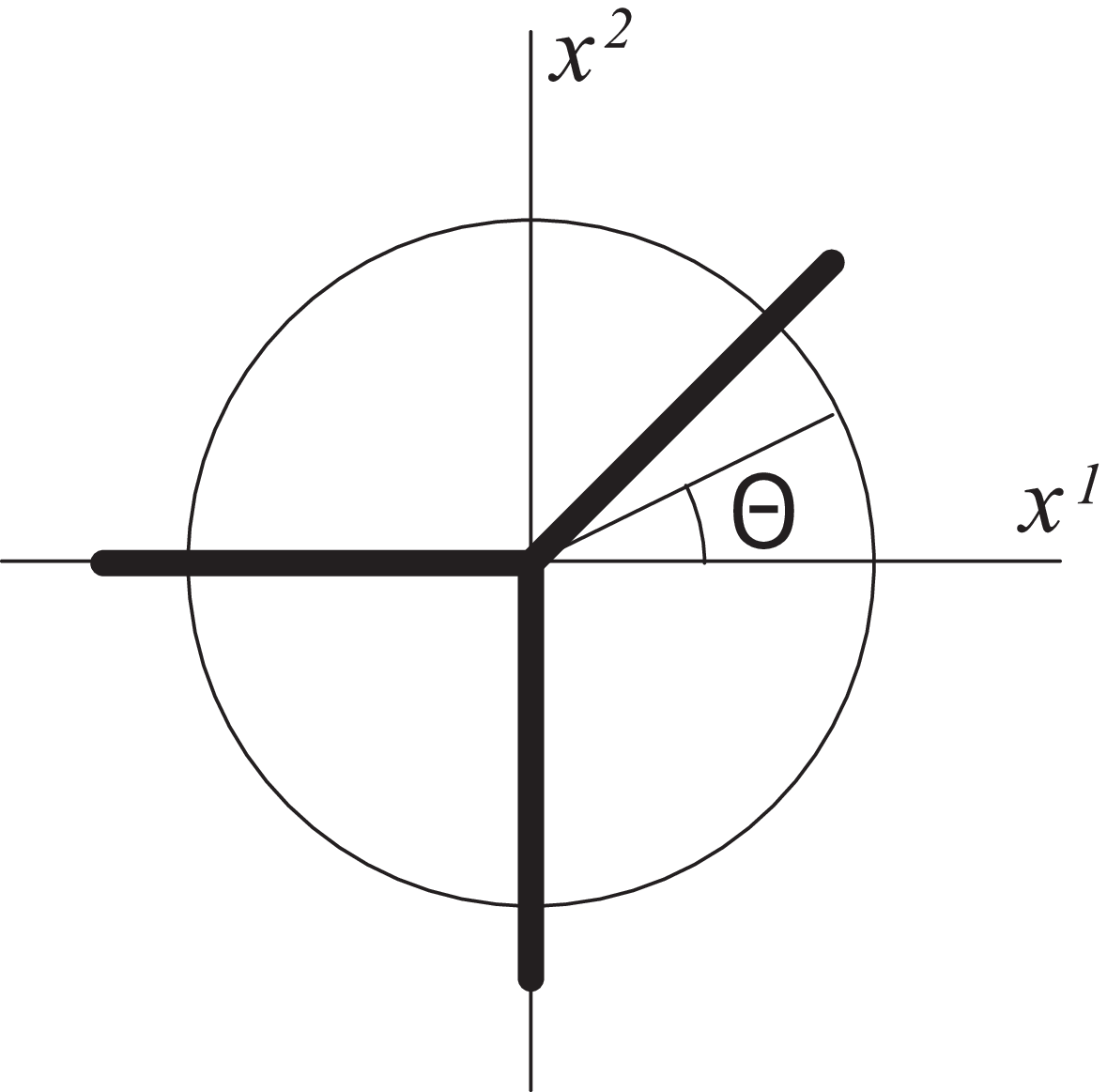} 
 &  \includegraphics[width=3cm,clip]{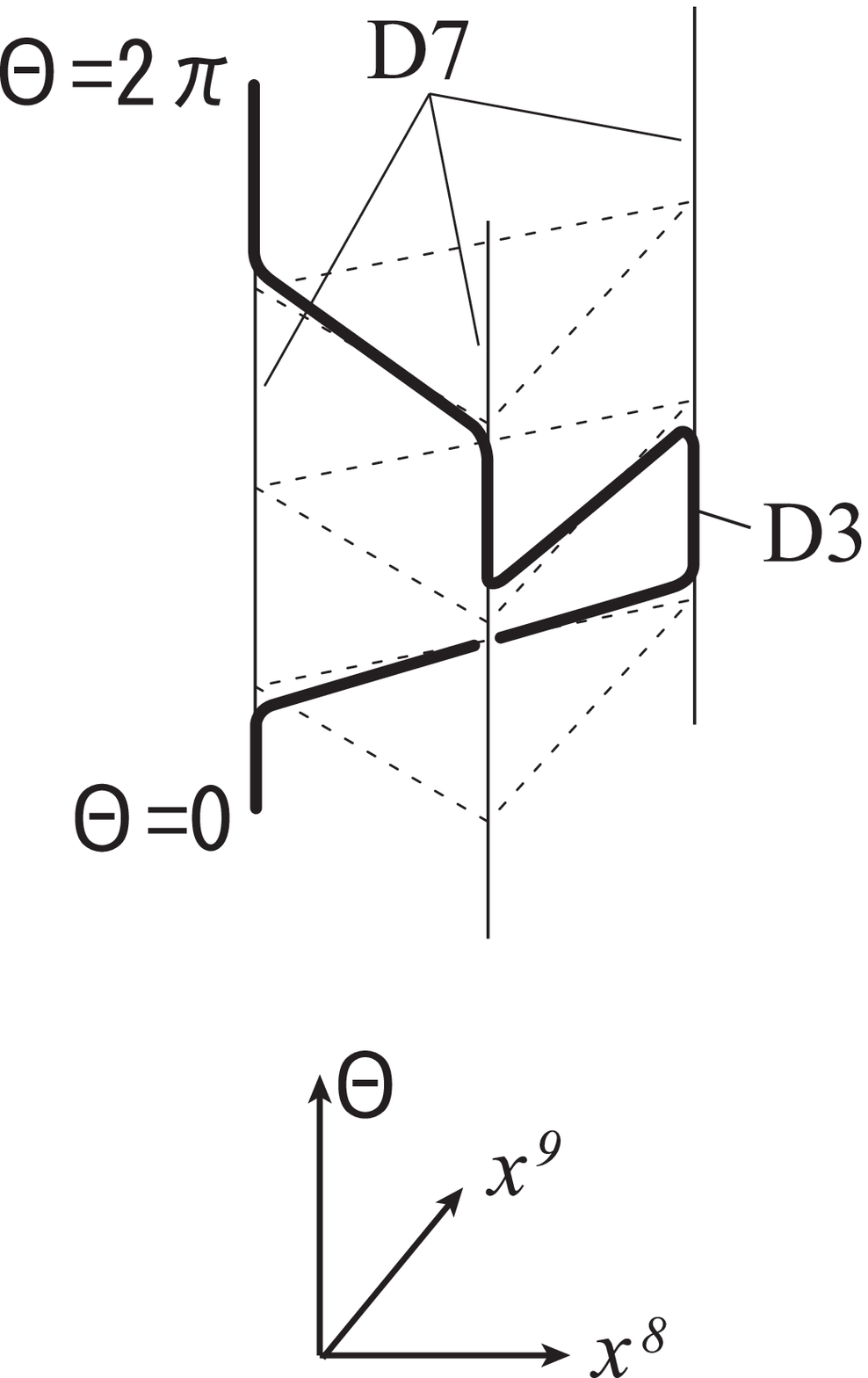} \\
  a) A three wall junction &  b) Brane configuration
\end{tabular} 
\caption{\small
a) Three domain wall junction in the real space $x^1$ and $x^2$. 
The origin is placed at the junction point. 
We take a polar coordinate $x^1 = r \cos \theta$ and 
$x^2 = r \sin \theta$. 
b) Brane configuration for the three wall junction in 
$U(1)$ gauge theory with $\Nf=3$. 
Three flavors with complex masses 
correspond to three D$7$-branes in the $x^8$-$x^9$ plane.  
One D$3$-brane interpolate between these three D$7$-branes. 
With fixing $r$ sufficiently large, 
the map from the angle $\theta$ to 
$(x^8, x^9 ) = l_s^2 (\Sigma^1 (r,\theta),\Sigma^2 (r,\theta))$ 
is illustrated. 
The D$3$-brane does not touch 
any D$7$-branes for smaller $r$, 
and for all $r$ it sweeps inside the triangular column 
made of the three D$7$-branes.
}
\label{fig-juncy}
\end{figure}

Several walls form a 1/4 BPS wall junction when 
both $\Sigma^1$ and $\Sigma^2$ fully 
depend on both $x^1$ and $x^2$ 
as a solution of (\ref{bps_eq1}). 
So the configurations are given by maps 
from $(x^1,x^2)$ to $(x^8,x^9)$. 
An example of three wall junction in the $U(1)$ gauge theory 
with $\Nf=3$ is illustrated in Fig.~\ref{fig-juncy}.
Along a circle with sufficiently large radius 
surrounding an $n$-wall junction point in the real space, 
we encounter $n$ vacua. 
That circle is mapped to the brane configuration 
with D$3$-branes successively interpolating 
between $n$ D$7$-branes. 
For smaller radius the D$3$-brane is detached from 
the D$7$-branes and 
goes inside the region sorrounded by $n$ D$7$-branes. 
Varying $r$ the configuration sweeps inside that column.

\begin{figure}[thb]
\begin{tabular}{cc}
 \includegraphics[width=4cm,clip]{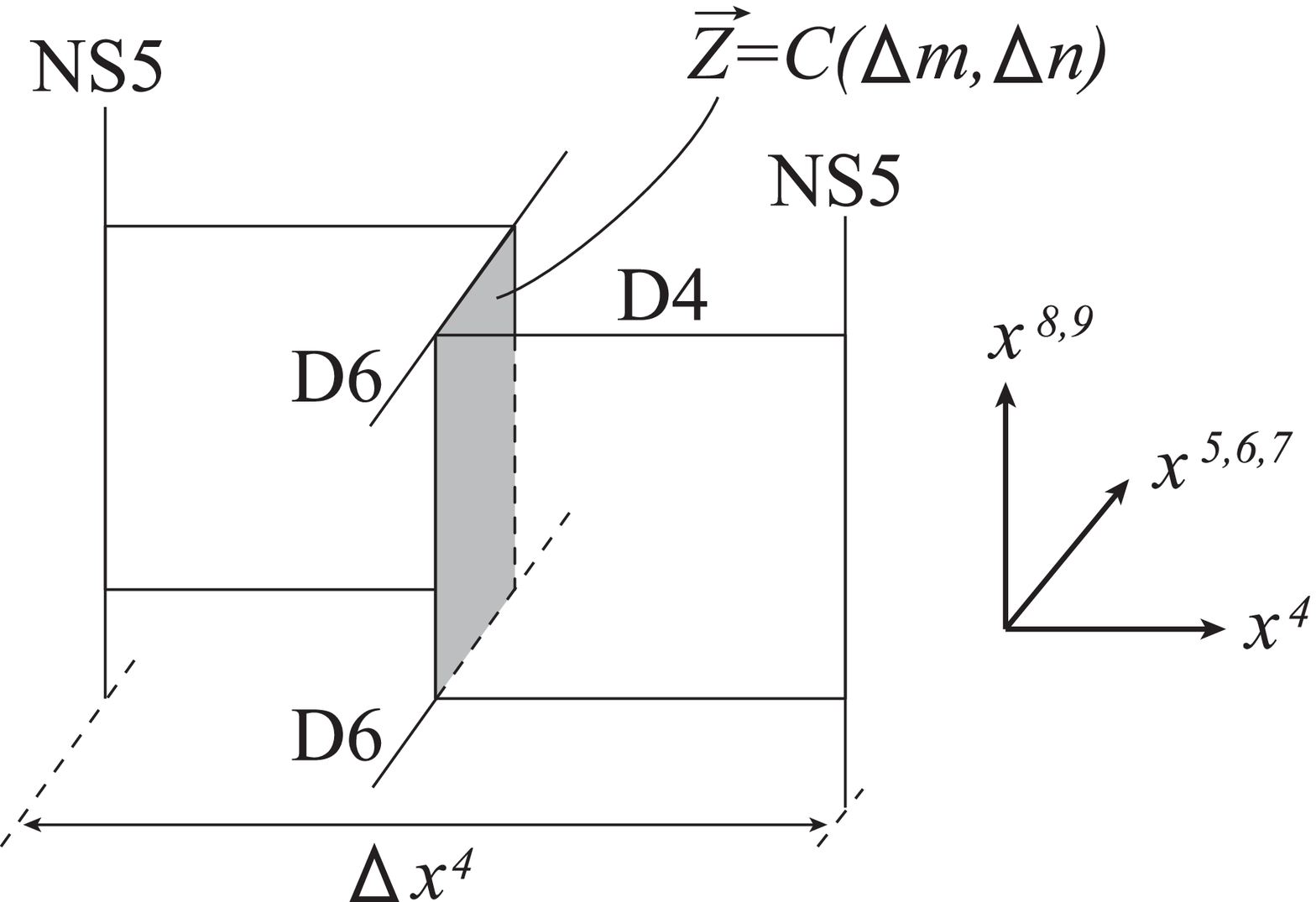} 
 &  \includegraphics[width=3cm,clip]{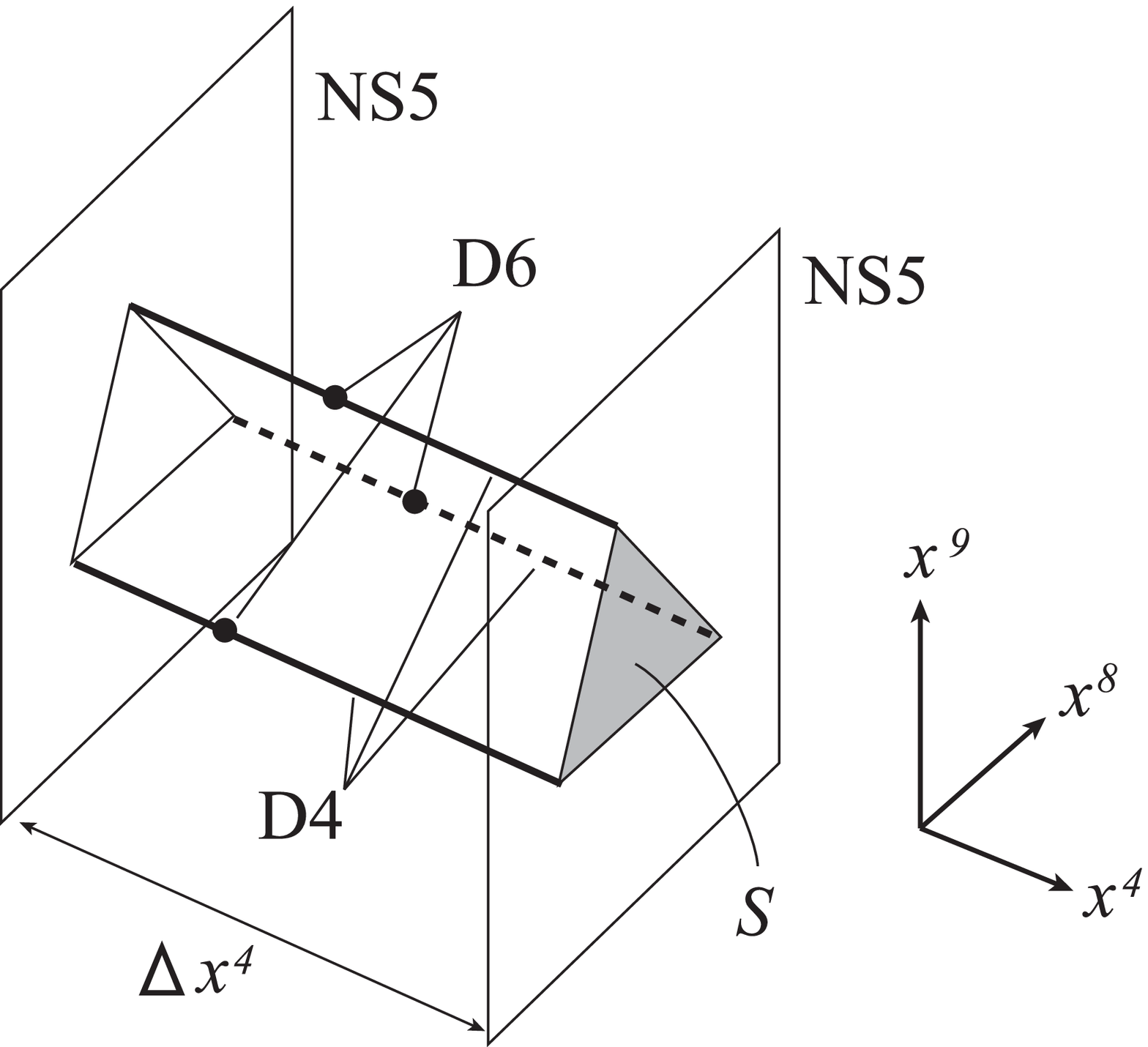} \\
  a) Wall charges $Z_{\alpha}$  & b) Junction charge $Y$
\end{tabular} 
\caption{\small
Brane configuration for domain wall junction.
}
\label{fig-d6d4}
\end{figure}

Let us take a T-duality 
along $x^4$ as
(\ref{HWset1}) and (\ref{quantities}). 
The brane configuration for wall junction 
is illustrated in Fig.~\ref{fig-d6d4}.
In these figures, 
the dependence of $\Sigma$ to $x^1$ and $x^2$ is 
suppressed but is drawn in the same figure. 
Each D$4$-brane ends on a D$6$-brane at one spatial infinity of 
the $x^1$-$x^2$ plane with fixing $x^1/x^2$. 
It can move to another D$6$-brane at another spatial infinity. 
Then the position $\Sigma_{1,2}$ 
of a D$4$-brane at each point of the $x^1$-$x^2$ plane 
is drawn in the same figure. 
The D$4$-brane tension contributing to 
$\tau_{4}\Delta x^4 (\Delta x^8, \Delta x^9 ) 
 = \1{g^2 l_s^2} (\Delta m , \Delta n )$
diverges in the limit $l_s \to 0$ but
this is the vacuum energy.
The contribution to the shaded area in Fig.~\ref{fig-d6d4}-a) 
\beq
 \tau_{4} \Delta x^7 (\Delta x^8, \Delta x^9 ) 
 = c (\Delta m, \Delta n) = (Z_1, Z_2)
\eeq
recovers the wall tensions correctly. 
We can only see the wall in Fig.~\ref{fig-d6d4}-a) 
but not a junction. 
To see a wall junction and its junction charge $Y$,  
we ignore the dependence on $x^5,x^6,x^7$. 
Instead in Fig.~\ref{fig-d6d4}-b), 
we take into account the $x^8$-$x^9$ dependence. 
Just as in Fig.~\ref{fig-d6d4}-a), 
we draw D$4$-brane position at each ($x^1,x^2$) 
in the same figure.  
Then it sweeps inside a triangular column. 
Interestingly the junction charge 
$Y$ can be calculated 
from the product of the D$3$-brane tension and 
the volume of the triangular column
\beq
 Y \sim \tau_4 \mbox{Vol (triang. \hs{0.5} column)} 
   =  \tau_4 \Delta x^4 |dx^8 \wedge dx^9|. 
\eeq 
$Y$ is actually negative for $U(1)$ gauge theory 
\cite{webs} but it is not clear 
in the brane picture yet. 
Non-Abelian $U(\Nc)$ case could be considered 
in the same way. 
As in domain wall case \cite{Eto:2004vy} 
duality between junctions in 
$U(\Nc)$ and  $U(\Nf-\Nc)$ gauge theories 
would be explained by exchange of 
the two NS$5$-branes. 
%

\section{Discussion} \label{CD}
In this talk we have discussed 
only non-degenerate masses for hypermultiplets. 
When masses are degenerated 
more interesting physics 
appears \cite{Shifman:2003uh,Eto:2005cc}. 
Other than domain walls, 
lots of composite BPS solitons in the Higgs phase 
have been found, like 
1/4 BPS instantons with vortices \cite{Eto:2004rz}, 
1/4 BPS intersecting vortices \cite{Naganuma:2001pu}, 
and various 1/8 BPS composite solitons \cite{1/8}. 
Reviews of some of these development can be found in \cite{review}. 
D-brane configurations for these newly found 
solitons are desired.


\begin{theacknowledgments}

This work is supported in part by Grant-in-Aid for Scientific 
Research from the Ministry of Education, Culture, Sports, 
Science and Technology, Japan No.13640269 (NS) 
and 16028203 for the priority area ``origin of mass'' 
(NS). 
M.~N. and K.~Ohashi. (M.~E. and Y.~I.) are 
supported by Japan Society for the Promotion 
of Science under the Post-doctoral (Pre-doctoral) Research 
Program while 
K.~Ohta is supported in part by Special Postdoctoral Researchers
Program at RIKEN.

\end{theacknowledgments}

\end{document}